\documentclass[%
reprint,twocolumn,
amsmath,amssymb,
aps,
pra,
]{revtex4-2}

\usepackage{graphicx}
\usepackage{epstopdf}
\usepackage{dcolumn}
\usepackage{bm}
\usepackage{amsfonts,amsmath,amssymb
}

\usepackage{bm,dsfont}
\usepackage{color}
\usepackage{bbm}
\usepackage{longtable}
\usepackage[dvips]{epsfig}
\usepackage{hyperref}
\usepackage{listings}
\usepackage{changes} 
\usepackage{todonotes}

\newcommand{\bra}[1]{\langle #1|}

\newcommand{\ket}[1]{|#1 \rangle}

\newcommand{\braket}[2]{\langle #1|#2 \rangle}

\renewcommand{\vec}[1]{\ensuremath{\boldsymbol #1 }}

\begin{document}
	\title{Conversion from $W$ to Greenberger-Horne-Zeilinger states
               in the Rydberg-blockade regime of neutral-atom systems: 
	       Dynamical-symmetry-based approach}
	
	\author{Thorsten Haase, Gernot Alber, and Vladimir M. Stojanovi\'c}
	\affiliation{Institut f\"{u}r Angewandte Physik, Technical
		University of Darmstadt, D-64289 Darmstadt, Germany}
	\date{\today}
	\begin{abstract}
		We investigate the possibilities for a deterministic conversion between two important types of 
                maximally entangled multiqubit states, namely, $W$ and Greenberger-Horne-Zeilinger (GHZ) states,
                in the Rydberg-blockade regime of a neutral-atom system where each atom is subject to four external 
                laser pulses. Such interconversions between $W$ states and their GHZ counterparts have quite recently 
                been addressed using the method of shortcuts to adiabaticity, more precisely techniques based on 
                Lewis-Riesenfeld invariants [R.-H. Zheng {\em et al.}, Phys. Rev. A {\bf 101}, 012345 (2020)]. 
                Motivated in part by this recent work, we revisit the $W$ to GHZ state-conversion problem using a 
                fundamentally different approach, which is based on the dynamical symmetries of the system and a
                Lie-algebraic parametrization of its permissible evolutions. In contrast to the previously used 
                invariant-based approach, which leads to a state-conversion protocol characterized by strongly 
                time-dependent Rabi frequencies of external lasers, ours can also yield one with time-independent 
                Rabi frequencies. This feature makes our protocol more easily applicable experimentally, with the 
                added advantage that it allows the desired state conversion to be carried out in a significantly 
                shorter time with the same total laser pulse energy used.
	\end{abstract}
	
	\maketitle
\section{Introduction}
Recent years have witnessed considerable progress in quantum-state engineering in a variety of physical systems~\cite{Haas+:14,
Song+:17,Friis+:18}, this being one of the essential prerequisites for the development of future quantum technologies~\cite{Dowling+Milburn:03,
Schleich+:16}. In particular, the recent milestones pertaining to the scalability of neutral-atom arrays trapped in optical 
tweezers~\cite{Barredo+:16,Bernien+:17,Barredo+:18,Brown+:19,deMello+:19,Schymik+:20} have reinvigorated interest in quantum-state 
engineering in ensembles of Rydberg atoms~\cite{GallagherBOOK,Adams+:20}. These 
developments~\cite{Buchmann+:17,Ostmann+:17,Omran+:19,Zheng+:20,Mukherjee+:20,Stojanovic:21}, in turn, bode well for future 
progress in the realm of 
quantum-information processing (QIP) with this class of atomic systems~\cite{Theis+:16,Levine+:18}, a research direction whose overarching 
goal is the realization of a neutral-atom quantum computer~\cite{Saffman+MoelmerRMP:10,Saffman:16}. High-fidelity state preparation/readout, 
quantum logic gates, and controlled quantum dynamics of more than $50$ qubits have already been demonstrated in those systems, with the 
prospect of reliable QIP with hundreds of qubits being within reach~\cite{Morgado+Whitlock:20}.

Maximally-entangled multi-qubit states are of particular importance for QIP, regardless
of the specific physical platform. Prominent ones among them are $W$~\cite{Duer+:00} and Greenberger-Horne-Zeilinger 
(GHZ)~\cite{Greenberger+Horne+Zeilinger:89} states, two classes of states that cannot be transformed into each other through local operations 
and classical communication (LOCC-inequivalence~\cite{Nielsen:99}). Owing to their favorable properties, both classes have 
proven useful in various QIP protocols~\cite{Joo+:03,Zhu+:15,Lipinska+:18}, which triggered a large number of proposals 
for the preparation of $W$~\cite{Tashima+:08,Peng+:0910,Gangat+:13,Li+Song:15,Kang+:16,Kang+SciRep:16,Fang+:19,StojanovicPRL:20,Bugu+:20} 
and GHZ states~\cite{Coelho+:09,Song+:17,Erhard+:18,Macri+:18,Zheng+:19} in various systems. 

Aside from tailored schemes for the preparation of $W$ or GHZ states, which typically involve a simple product state 
as their point of departure~\cite{Omran+:19}, the interconversion between a $W$ state and its GHZ counterpart represents 
another interesting problem of quantum-state engineering. The earliest attempt in this direction pertained to a photonic system and was 
probabilistic in nature~\cite{Walther+:05}. This pioneering study was followed by a further work with photons~\cite{Cui+:16} 
and a study of analogous interconversions in a spin system~\cite{Kang+:19}. Finally, irreversible conversions of a $W$ state into 
a GHZ state were also proposed in the realm of atomic systems~\cite{Song+:13,Wang+:16}. 
	
Quite recently, deterministic interconversions between $W$ and GHZ states have been investigated in the Rydberg-blockade (RB) 
regime~\cite{RBmanybody:04,RBobservation:09} of a neutral-atom system subject to four external laser pulses~\cite{Zheng+:20}. Each 
atom in this system was assumed to represent an effective two-level system, 
with the two relevant states (ground- and a Rydberg state) playing the roles of logical qubit states. This recent work was based 
on the method of shortcuts to adiabaticity (STA)~\cite{STA_RMP:19}. More specifically yet, it utilizes the concept of Lewis-Riesenfeld 
invariants~\cite{Lewis+Riesenfeld:69}, in this case applied to an effective four-level Hamiltonian of the system. RB, the phenomenon 
whereby the van der Waals (vdW) interaction prevents simultaneous Rydberg excitation of more than one atom within a certain radius, 
represents the enabling physical mechanism for QIP with neutral atoms~\cite{Saffman+MoelmerRMP:10} as it engenders a conditional logic 
that permits the realization of entangling two-qubit gates~\cite{Lukin+:01,Wilk+:10,Isenhower+:10}. From the point of view of quantum-state 
engineering, the most important implication of RB is that it leads to the creation of coherent superpositions with a single Rydberg 
excitation being shared among all atoms in an ensemble~\cite{Wilk+:10}. Such superpositions are maximally-entangled states of $W$ 
type~\cite{Saffman+Moelmer:09}. 
	
In this paper, we study the conversion of an initial $W$ state into a GHZ state in the same physical setting as Ref.~\cite{Zheng+:20} -- i.e., 
the RB regime of a system of neutral atoms interacting through vdW interactions. However, we employ a fundamentally different approach than 
that of Ref.~\cite{Zheng+:20}. Namely, our approach entails the use of the dynamical symmetries~\cite{BarutBOOK:71,Rau+Alber:17} of the system 
under consideration and is mathematically framed using the language of Lie algebras. Another important technical ingredient of our approach 
is a particular parametrization  of all the unitary 
transformations connecting arbitrary two states of the four-dimensional representation of the Lie algebra $so(4)$~\cite{RichtmyerBOOK2:81}. 
We utilize this parametrization to describe the dynamics corresponding to the effective system Hamiltonian that depends on the aforementioned 
Rabi frequencies. We then single out specific unitary evolutions of the system that connect the initial- and final states ($W$ and GHZ, 
respectively) in the quantum-state control problem at hand. This allows us to determine the corresponding Rabi frequencies of external laser
pulses.
	
In contrast to the STA-based approach of Ref.~\cite{Zheng+:20}, which results in a state-conversion protocol characterized by time-dependent 
Rabi frequencies with a rather complex time dependence, our approach can also yield one with constant (i.e., time-independent) Rabi frequencies. 
Importantly, this last feature makes our resulting protocol more easily applicable experimentally than that of Ref.~\cite{Zheng+:20}, with 
the added benefit that it allows the desired state conversion to be carried out in a significantly shorter time.

The remaining part of this work is organized in the following manner. In Sec.~\ref{System} we introduce the 
Rydberg-atom system under consideration, specifying at the same time the notation and conventions to be used 
throughout the paper. In Sec.~\ref{DynSymmApproach} we lay the groundwork for our dynamical-symmetry-based approach,
with emphasis on its Lie-algebraic description. Section~\ref{HamDyn} is concerned with the Hamiltonian dynamics, 
first in the most general case and then under the constraints inherent to the model used. In Sec.~\ref{WtoGHZ} 
we first discuss the specific unitary evolutions of the system that are required for carrying out the desired state 
conversion. We then determine the corresponding Rabi frequencies. Finally, we compare our resulting state-conversion 
protocol with the one obtained in Ref.~\cite{Zheng+:20}, with emphasis on the robustness of those protocols against 
decoherence effects. We close in Sec.~\ref{SummConcl} with a summary of the results and questions for future work. 
Some involved mathematical details are relegated to Appendices \ref{app:rotSTbasis} and \ref{app:condGHZ}.

\section{System and effective Hamiltonian} \label{System}
\begin{figure}[t!]
\includegraphics[width=8.6cm]{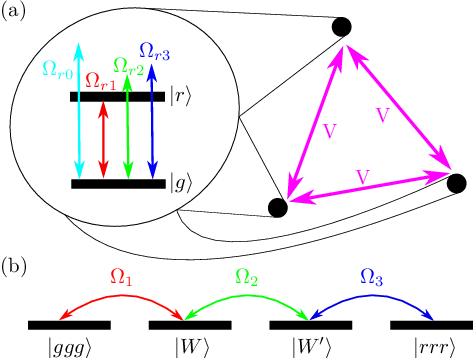} 
\caption{\label{fig:model_Heff}(Color online)(a) Schematic of the system under consideration: Three identical neutral atoms with 
internal states $\ket{g}$ and $\ket{r}$ ($gr$-type qubits) and constant vdW interaction energy $\hbar V$, which are subject to four 
external laser pulses. \\
(b) Pictorial illustration of the effective Hamiltonian $H_{\textrm{eff}}$ in Eq.~\eqref{eq:HeffZheng}.}
\end{figure}

We consider a system that consists of three equidistant and identical neutral atoms in the RB regime [for an illustration, see 
Fig.~\ref{fig:model_Heff}(a)]. All three atoms are subject to the same four external laser pulses, with their corresponding Rabi 
frequencies being denoted by $\Omega_{r0}$, $\Omega_{r1}$, $\Omega_{r2}$, and $\Omega_{r3}$. These pulses, with frequencies 
$\omega_i$ $(i \in \{0,1,2,3\})$, are close to being resonant with only a single internal (electronic) transition $\ket{g}\leftrightarrow\ket{r}$. 
Thus, each atom can effectively be treated as a two-level system with its electronic ground state $|g\rangle$ and a highly-excited 
Rydberg state $|r\rangle$. These states encode the logical $|0\rangle$ and $|1\rangle$ qubit states, respectively. 
Therefore, from the QIP standpoint this is a system of neutral-atom qubits of the ground-Rydberg ($gr$) type~\cite{Morgado+Whitlock:20}. 
Note that the typical energy splitting of such qubits in frequency units is $(900 - 1500)$\:THz (depending on the choice of atomic 
species and the Rydberg states used), thus in practice $gr$-qubit manipulations entail either an ultraviolet laser or a combination 
of visible and infrared lasers in a ladder configuration. Owing to their relatively straightforward initialization, manipulation, and 
measurement -- as compared to other kinds of neutral-atom qubits -- $gr$ qubits currently represent the preferred qubit type for fast 
($\lesssim 100$\:ns) and high-fidelity entangling operations~\cite{Levine+:18}, as well as for quantum-state engineering~\cite{Omran+:19}.

The interaction-picture Hamiltonian of the system is given by
\begin{align}
H_{\textrm{I}}(t) /\hbar =& \sum_{k=1}^3\sum_{i=0}^3\Omega_{ri}(t)e^{-i(\delta_i+\Delta_i) t}\ket{r}_{kk}\bra{g}
+ \text{H.c.} \nonumber \\
& + \sum_{p<q} V\ket{rr}_{pq}\bra{rr} \ . \label{eq:Hint}
\end{align}
Here, the laser frequencies are slightly detuned from the atomic transition frequency, i.e. $\omega_i = (E_r-E_g)/\hbar+\Delta_i+\delta_i$ 
with $i\in\{0,1,2,3\}$. The Rabi frequencies $\Omega_{r1}$, $\Omega_{r2}$, and $\Omega_{r3}$ are time dependent and real-valued, while 
$\Omega_{r0}$ is time independent and plays the role of inducing appropriate quadratic Stark shifts. $\hbar V$ is the constant vdW interatomic 
interaction energy. The interatomic interaction between each pair of excited Rydberg atoms is described approximately by an energy shift 
of magnitude $\hbar V$ according to lowest order perturbation theory.

As shown in Ref.~\cite{Zheng+:20}, by choosing the detunings $\delta_i$ and $\Delta_i$ appropriately, the system Hamiltonian can be reduced 
via perturbation theory to an effective Hamiltonian defined in the manifold of four three-particle states $\ket{ggg}$, $\ket{W} = (\ket{rgg}+\ket{grg}+
\ket{ggr})/\sqrt{3}$, $\ket{W'}= (\ket{rrg}+\ket{grr}+\ket{rgr})/\sqrt{3}$, and $\ket{rrr}$. To be more specific, choosing the detunings 
such that $\delta_0=0$, $\Delta_1 = 0$, $\Delta_2 = V$, $\Delta_3 = 2V$, and
\begin{align}
\delta_1 =& -\frac{6\Omega_{r0}^2}{\Delta_0}+\frac{4\Omega_{r0}^2}{\Delta_0-V} \:,\nonumber \\
\delta_2 =& \frac{3\Omega_{r0}^2}{\Delta_0}-\frac{8\Omega_{r0}^2}{\Delta_0-V}+\frac{3\Omega_{r0}^2}{\Delta_0-2V} \:,\nonumber \\ 
\delta_3 =& \frac{4\Omega_{r0}^2}{\Delta_0-V}-\frac{6\Omega_{r0}^2}{\Delta_0-2V} \:,
\end{align}
yields the effective Hamiltonian
\begin{align}
H_{\textrm{eff}}(t)/\hbar =& \Omega_1(t)\ket{ggg}\bra{W} +\Omega_{2}(t)\ket{W}\bra{W'}\nonumber\\ &+\Omega_{3}(t)
\ket{W'}\bra{rrr} +\text{H.c.} \:,\label{eq:HeffZheng}
\end{align}
where $\Omega_1(t)=\sqrt{3}\:\Omega_{r1}(t)$, $\Omega_2(t)=2\Omega_{r2}(t)$, and $\Omega_3(t)=\sqrt{3}\Omega_{r3}(t)$. 

In addition to the two-level approximation a sufficient set of conditions for the validity of this effective Hamiltonian requires 
sufficiently long interaction times $T_{int}$, i.e. ${\rm min}\{|\Delta_0|, |V| \} T_{int} \gg 1$,
and $|\Delta_0|, |V|\gg |\Omega_{r0}|, {\rm min}\{| \delta_i |; i=1,2,3 \} \gg {\rm max}\{ |\Omega_{ri}(t)|; i=1,2,3 \}$.

It is pertinent to comment at this point on the realm of applicability of the last effective 
Hamiltonian -- as well as the ensuing state-conversion scheme -- vis-\`{a}-vis the RB regime, as the latter represents 
the regime of primary interest for QIP with neutral atoms~\cite{Saffman+MoelmerRMP:10}. In this context it is useful 
to recall that the RB regime is defined as the one in which the interaction-induced energy shift $\hbar V$ is much larger 
than the Fourier-limited width of the laser pulses involved (i.e. $\mid V\mid T_{\textrm{int}} \gg 1$, where $T_{\textrm{int}}$
is the pulse duration). Because this last condition is satisfied for all the laser pulses envisioned to be used in our 
present state-conversion scheme, we can identify the RB regime as the domain of applicability of our approach.

In the following it is demonstrated that just on the basis of the $so(4)$ dynamical symmetry of the last effective Hamiltonian 
alone, i.e. without invoking the Lewis-Riesenfeld method of Ref.~\cite{Zheng+:20}, a simpler and more time-efficient protocol for 
the desired conversion of $W$ states into their GHZ counterparts can be found.

\section{Dynamical Symmetries} \label{DynSymmApproach}
In what follows the $su(2)\oplus su(2) \cong so(4)$ dynamical symmetry of the effective Hamiltonian in Eq.~\eqref{eq:HeffZheng} 
is investigated. It is shown that under the assumption of real-valued Rabi frequencies $\Omega_i(t)$ ($i=1,2,3$) this last Hamiltonian 
describes the quantum dynamics of two constrained pseudospin-$1/2$ degrees of freedom.

To begin with, we map the orthonormal basis states in the Hamiltonian of Eq.~\eqref{eq:HeffZheng} onto column vectors according to
\begin{eqnarray}
\ket{ggg}&\rightarrow&
\left(\begin{array}{c}
1\\
0\\
0\\
0
\end{array}\right),~~\ket{W} ~\rightarrow~
\left(\begin{array}{c}
0\\
1\\
0\\
0
\end{array}\right),\nonumber\\
\ket{W'} &\rightarrow&
\left(\begin{array}{c}
0\\
0\\
1\\
0
\end{array}\right),~~\ket{rrr}~\rightarrow~
\left(\begin{array}{c}
0\\
0\\
0\\
1
\end{array}\right).
\label{physicalstates}
\end{eqnarray}
Within this notation the six matrices
\begin{align}\label{eq:Liealgebrarepresentation}
S_1 &= \frac{1}{2}\begin{pmatrix}0 & 1 & 0 & 0 \\ 1 & 0 & 0 & 0\\ 0 & 0 & 0 & 1 \\ 0 & 0 & 1 & 0 \end{pmatrix} ,& 
T_1 &= \frac{1}{2}\begin{pmatrix}0 & 1 & 0 & 0 \\ 1 & 0 & 0 & 0\\ 0 & 0 & 0 & -1 \\ 0 & 0 & -1 & 0\end{pmatrix}, \nonumber \\
S_2 &= \frac{1}{2}\begin{pmatrix}0 & 0 & 0 & -1 \\ 0 & 0 & 1 & 0\\ 0 & 1 & 0 & 0 \\ -1 & 0 & 0 & 0 \end{pmatrix} ,&  
T_2 &= \frac{1}{2}\begin{pmatrix}0 & 0 & 0 & 1 \\ 0 & 0 & 1 & 0\\ 0 & 1 & 0 & 0 \\ 1 & 0 & 0 & 0 \end{pmatrix},\\
S_3 &= \frac{1}{2}\begin{pmatrix}0 & 0 & -i & 0 \\ 0 & 0 & 0 & i\\ i & 0 & 0 & 0 \\ 0 & -i & 0 & 0 \end{pmatrix} , &  
T_3 &= \frac{1}{2}\begin{pmatrix}0 & 0 & -i & 0 \\ 0 & 0 & 0 & -i\\ i & 0 & 0 & 0 \\ 0 & i & 0 & 0 \end{pmatrix} \nonumber ,
\end{align}
with Lie brackets
\begin{eqnarray}\label{Liebrackets}
\left[S_i,S_j\right]&=&i\epsilon_{ijk}S_k \:, \nonumber\\
\left[T_i,T_j\right]&=& i\epsilon_{ijk}T_k\:, \\
\left[S_i,T_j\right]&=& 0 \:, \nonumber
\end{eqnarray}
constitute a four dimensional representation of the Lie algebra $su(2)\oplus su(2)\cong so(4)$ and thus
describe two independent angular momenta. From the relevant Casimir operators
\begin{equation}
I = \sum_{i=1}^3 \left(S_i^2+T_i^2\right)= \frac{3}{2}\:, \: J=\sum_{i=1}^3\left(S_i^2-T_i^2\right)=0 
\end{equation}
it is apparent that the (dimensionless) two angular momentum operators with Cartesian components 
$\{S_i\}$ and $\{T_i\}$ ($i\in \{1,2,3\}$) describe two independent spin-$1/2$ degrees of freedom or pseudospins 
because $\sum_{i}S_i^2=\sum_{i}T_i^2= s(s+1)$ with $s=1/2$. Furthermore, the matrices of the representation 
[cf. Eq.~\eqref{eq:Liealgebrarepresentation}] fulfill the additional relations
\begin{eqnarray}
(2S_l)(2S_k) &=&i\epsilon_{lkm}(2S_m) + \delta_{lk},~~l,k,m\in \{1,2,3\}\nonumber\\
(2T_l)(2T_k) &=&i\epsilon_{lkm}(2T_m) + \delta_{lk}
\end{eqnarray}
in analogy to two commuting sets of Pauli spin matrices.

Provided the Rabi frequencies $\Omega_i(t)$ are real-valued the Hamiltonian of Eq.~\ref{eq:HeffZheng}
is a linear combination of these angular momentum operators, i.e.
\begin{eqnarray}
H_{\textrm{eff}}(t)/\hbar &=& \Omega_1(t) \left(S_1 + T_1 \right) + \Omega_2(t) \left(S_2 + T_2 \right) \nonumber\\
&& +\Omega_3(t) \left(S_1 - T_1 \right) .
\label{EffHam1}
\end{eqnarray}
This explicitly exhibits the Lie algebra $su(2)\oplus su(2)$ as the dynamical symmetry of this Hamiltonian~\cite{BarutBOOK:71,Rau+Alber:17}. 
However, it is also apparent that this effective Hamiltonian is not the most general real-valued linear combination of the six independent  
operators $S_i$ and $T_i$ ($i\in \{1,2,3\}$). It is constrained by the fact that some operators, such $S_2 - T_2$, do not appear in this Hamiltonian.

In view of the $su(2)\oplus su(2)$ symmetry of the Hamiltonian \eqref{eq:HeffZheng} it is convenient to use the eigenstates of 
the spin operators of the two independent pseudospins as an orthonormal basis of the Hilbert space. These states can straightforwardly 
be constructed by starting from the common eigenvector of the operator $S_3 + T_3$ with the largest eigenvalue, i.e. 
\begin{eqnarray}
\ket{\uparrow\uparrow} &\rightarrow& \frac{1}{\sqrt{2}}
\left(\begin{array} {c}
-i \\ 0 \\ 1 \\ 0
\end{array}\right)
\label{upup}
\end{eqnarray}
with $(S_3 + T_3)\ket{\uparrow \uparrow} = \ket{\uparrow \uparrow}$. By applying the lowering operators $S_1-iS_2$ and $T_1-iT_2$ 
onto $\ket{\uparrow\uparrow}$ the remaining orthonormal states of the independent pseudospins are obtained, namely
\begin{eqnarray}\label{eq:STbasis} 
\ket{\uparrow\downarrow}&\rightarrow& \frac{1}{\sqrt{2}}
\left(\begin{array}{c}  0 \\ -i \\ 0 \\ -1\end{array}\right),~~
\ket{\downarrow\uparrow}~\rightarrow ~\frac{1}{\sqrt{2}}
\left(\begin{array}{c} 0 \\ -i \\ 0 \\ 1\end{array}\right),\nonumber\\\ket{\downarrow\downarrow}&\rightarrow& 
\frac{1}{\sqrt{2}}\left(\begin{array}{c}  -i \\ 0 \\ -1 \\ 0\end{array}\right)\:. 
\label{uprest}
\end{eqnarray} 

In this basis the angular momentum operators $S_i$ ($T_i$) act on the first (second) pseudospin only. Therefore, these orthonormal 
basis states transform under an arbitrary unitary transformation in a simple way. In view of $\left[ S_i,T_j \right]=0,~i,j\in \{1,2,3\}$ 
the most general unitary transformation generated by these commuting angular momenta can be written in the form
\begin{eqnarray}
U(\vec{\alpha},\vec{\beta}) &=& e^{-i\vec{\alpha}\cdot\vec{S}}
e^{-i\vec{\beta}\cdot\vec{T}}.
\label{unitary}
\end{eqnarray} 
Explicit expressions for the transformed pseudospin states of Eqs.~\eqref{upup} and \eqref{uprest} are given in 
Appendix~\ref{app:rotSTbasis}.

\section{Unitary transformations and time-dependent Hamiltonians} \label{HamDyn}
In this section basic properties of time-dependent unitary operators are discussed
which are induced by time-dependent curves in the parameter space of the Lie algebra $su(2)\oplus su(2)$.  
Via the time-dependent Schr\"odinger equation their local properties are characterized by corresponding 
time-dependent Hamiltonians. 

Let us consider a general time-dependent unitary transformation $U[\vec{\alpha}(t),\vec{\beta}(t)]$
as defined by Eq.~\eqref{unitary} which is induced by a time-dependent (differentiable) curve, say $\gamma: 
t\longrightarrow\{\vec{\alpha}(t),\vec{\beta}(t)\}$ with $t\in \left[0,T\right]$, in the parameter space of the Lie 
algebra $su(2)\oplus su(2)$. Via the time-dependent Schr\"odinger equation such a time-dependent unitary transformation 
defines the corresponding time-dependent Hamiltonian $H(t)$, i.e.
\begin{eqnarray} 
i\hbar\frac{d}{dt}\:U[\vec{\alpha}(t), \vec{\beta}(t)] &=& H(t) U[\vec{\alpha}(t), \vec{\beta}(t)]\:.
\label{eq:Schroedinger}
\end{eqnarray}
This Hamiltonian $H(t)$ characterizes the local properties of this time evolution and is determined completely 
by the $su(2)\oplus su(2)$ commutation relations of Eq.~\eqref{Liebrackets}. It can be determined in a straightforward 
way from Eq.~\eqref{Uexplicit} of Appendix \ref{app:rotSTbasis}, thus yielding the result
\begin{eqnarray}
H(t)/\hbar &=&\vec{\omega}[\vec{\alpha}(t)]\cdot \vec{S} + \vec{\omega}[\vec{\beta}(t)]\cdot \vec{T} 
\label{unitary-Hamilton}
\end{eqnarray}
with the time-dependent (vectorial) Rabi frequency
\begin{eqnarray}
\vec{\omega}[\vec{\alpha}(t)] &=&
\frac{\sin|\vec\alpha(t)|}{|\vec{\alpha}(t)|}\:\dot{\vec{\alpha}}(t)+\frac{2\sin^2\frac{|\vec\alpha(t)|}{2}}
{|\vec\alpha(t)|^2}\:[\vec{\alpha}(t)\times\dot{\vec{\alpha}}(t)] \nonumber \\
&& +\frac{|\vec\alpha(t)|-\sin|\vec{\alpha}(t)|}{|\vec{\alpha}(t)|^3}\:[\vec{\alpha}(t)\cdot
\dot{\vec{\alpha}}(t)]\vec{\alpha}(t) \label{eq:generalHcoeff}
\end{eqnarray}
and with $\vec{\omega}[\vec{\beta}(t)]$ defined in an analogous fashion. [Note that here $\dot{\vec{\alpha}}(t)$ denotes 
the time derivative of $\vec{\alpha}(t)$.]

According to Eq.~\eqref{unitary-Hamilton} each 
time dependent (differentiable) curve $\gamma$
induces a time-dependent curve in the space of unitary transformations $U[\vec{\alpha}(t), \vec{\beta}(t)]$ whose associated 
time-dependent Hamiltonian $H(t)$ has the Lie algebra $su(2)\oplus su(2)$ as its dynamical symmetry. In general this 
time-dependent Hamiltonian $H(t)$ is a linear combination of all angular momenta of this Lie algebra. In the Hilbert space of the two 
pseudospins any pure quantum state, say $|\psi(0)\rangle$, can be converted into any other pure quantum state, say $|\psi(T)\rangle$, 
by the time-dependent Hamiltonian $H(t)$ of Eq.~\eqref{unitary-Hamilton} provided a curve $\gamma$ can be found with
\begin{eqnarray}
U[\vec{\alpha}(0), \vec{\beta}(0)]\ket{\psi(0)} &=& e^{i\Phi_0}\ket{\psi(0)},\nonumber\\
U[\vec{\alpha}(T), \vec{\beta}(T)]\ket{\psi(0)} &=& e^{i\Phi_T}\ket{\psi(T)}.
\label{conversion}
\end{eqnarray}
[Note that nonzero phases $\Phi_0, \Phi_T$ are possible in Eq.~\eqref{conversion} because pure quantum states are described 
by rays in Hilbert space.] If the form of the time-dependent Rabi frequencies of Eq.~\eqref{eq:generalHcoeff} are constrained by additional 
boundary conditions the possible curves $\gamma$ in the parameter space of the Lie algebra $su(2)\oplus su(2)$ may be restricted so that 
particular quantum state conversions are no longer possible. 

A typical example of a Hamiltonian with additional constraints is the effective time-dependent Hamiltonian of Eq.~\eqref{eq:HeffZheng}. 
Comparing Eqs.~\eqref{EffHam1} and \eqref{unitary-Hamilton} these constraints are explicitly given by
\begin{eqnarray}
\omega_3[\vec{\alpha}(t)] &=&0,\nonumber\\
\omega_3[\vec{\beta}(t)] &=&0,\nonumber\\
\omega_2[\vec{\alpha}(t)] - \omega_2[\vec{\beta}(t)]&=&0
\label{anhol}
\end{eqnarray}
and by the following 
relations
\begin{eqnarray}
\Omega_1(t) &=&\frac{\omega_1[\vec{\alpha}(t)] + \omega_1[\vec{\beta}(t)]}{2},\nonumber\\
\Omega_3(t) &=&\frac{\omega_1[\vec{\alpha}(t)] - \omega_1[\vec{\beta}(t)]}{2},\nonumber\\
\Omega_2(t) &=&\frac{\omega_2[\vec{\alpha}(t)] + \omega_2[\vec{\beta}(t)]}{2}
\label{relations2}
\end{eqnarray}
which relate the components of the (vectorial) Rabi frequencies to the time-dependent parameters of Eq.~\eqref{EffHam1}.
In particular, the relations \eqref{anhol} are anholonomic constraints that have to be fulfilled by all curves in the 
parameter space of the Lie algebra $su(2)\oplus su(2)$ which enable a pure state conversion with the aid of the effective 
Hamiltonian \eqref{eq:HeffZheng}.

\section{$W$-to-GHZ quantum state conversion} \label{WtoGHZ}
In this section we address the question as to which effective state conversions between a $W$ and a GHZ state are possible 
by appropriate choices of the time-dependent Rabi frequencies in the effective Hamiltonian of Eq.~\eqref{eq:HeffZheng}. In 
particular, it is demonstrated below that the $su(2)\oplus su(2)$-based Lie-algebraic approach allows one to determine quantum 
state conversions which are significantly faster than the previously proposed STA-based protocol~\cite{Zheng+:20}. 

We are looking for time-dependent Rabi frequencies $\Omega_i(t)$ ($i=1,2,3$) of the Hamiltonian [cf. Eq.~\eqref{eq:HeffZheng}] 
that allow the conversion of an initially prepared $W$ state into a GHZ state as described by Eq.~\eqref{conversion} with
\begin{eqnarray}
 \ket{\psi(0)} &=& \ket{W}, \label{eq:psit}\\
 \ket{\psi(T)} &=&\ket{\textrm{GHZ}} = \left(\ket{ggg} + e^{i\varphi}\ket{rrr}\right)\frac{1}{\sqrt{2}},~~\varphi\in[0,2\pi)\nonumber.
\end{eqnarray}
For this purpose we have to find an appropriate curve $\gamma$ in the parameter space of the Lie algebra $su(2)\oplus su(2)$ 
as described in Sec.~\ref{HamDyn} whose induced unitary transformation $U[\vec{\alpha}(t), \vec{\beta}(t)]$ acts onto the initial state 
like the unit transformation at $t=0$ and yields the final GHZ state at $t=T$. As this state conversion has to be achieved by 
the effective Hamiltonian \eqref{eq:HeffZheng} the time dependent Rabi frequencies have to fulfill the constraints \eqref{anhol} 
and the relations \eqref{relations2}.

Let us first of all determine curves $\gamma$ in the parameter space of $su(2)\oplus su(2)$ which fulfill the constraints \eqref{anhol}. 
For the sake of simplicity we restrict our subsequent discussion to curves $\gamma$ which fulfill the additional requirement 
that $\mid \vec{\alpha}(t)\mid = \mid \vec{\beta}(t)\mid = \pi$. In addition, expressing the vectorial parameters $\vec{\alpha}(t)$ 
and $\vec{\beta(t)}$ in spherical coordinates, i.e. $\alpha_1(t) = \pi \sin\theta_{\alpha}(t) \cos\phi_{\alpha}(t)$, $\alpha_2(t) = 
\pi \sin\theta_{\alpha}(t) \sin\phi_{\alpha}(t)$, $\alpha_3(t) = \pi \cos\theta_{\alpha}(t)$ and analogously for $\vec{\beta}(t)$,
 the anholonomic constraints of Eq.~\eqref{anhol} simplify to the relations
\begin{eqnarray}
0 &=&\sin^2(\theta_\alpha(t)) \dot{\phi}_\alpha(t), \label{eq:sphericalanhol1}\\
0 &=&\sin^2(\theta_\beta(t))\dot{\phi}_\beta(t), \label{eq:sphericalanhol2}\\
 0&=&\dot{\theta}_{\alpha}(t)\cos\phi_{\alpha}(t) - \dot{\theta}_{\beta}(t)\cos\phi_{\beta}(t)
 \label{eq:sphericalanhol3}
\end{eqnarray}
with $0\leq \theta_{\alpha}(t), \theta_{\beta}(t) \leq \pi$, $0\leq \phi_{\alpha}(t), \phi_{\beta}(t) < 2\pi$. 
The constraints imposed by Eqs.~\eqref{eq:sphericalanhol1}
and \eqref{eq:sphericalanhol2} can be fulfilled by choosing $\dot{\phi}_{\alpha}(t) =\dot{\phi}_{\beta}(t) = 0$ so 
that the general solution of Eq.~\ref{eq:sphericalanhol3} is given by
\begin{eqnarray}
\theta_{\beta}(t) &=& \theta_{\beta}(0) + \frac{\cos\phi_{\alpha}(0)}{\cos\phi_{\beta}(0)}\int_0^T dt~\dot{\theta}_{\alpha}(t).
\label{anholexplicit}
\end{eqnarray}
Apart from being integrable, the function $\dot{\theta}_{\alpha}(t)$ can be chosen arbitrarily.
According to Eqs.~\eqref{relations2}, the time dependencies of the Rabi frequencies are given by
\begin{eqnarray}
\Omega_1(t) &=&-\dot{\theta}_\alpha(t) 
\left(\sin \phi_\alpha+\cos\phi_{\alpha}\tan\phi_{\beta}\right) \nonumber \\
\Omega_2(t) &=&2\dot{\theta}_\alpha(t) \cos \phi_\alpha \nonumber \\
\Omega_3(t) &=&-\dot{\theta}_\alpha(t) \left(\sin \phi_\alpha-\cos\phi_{\alpha}\tan \phi_{\beta}\right)\:. \label{eq:omegapi}
\end{eqnarray}

As the next step, we proceed to determine the initial and final values of the curve $\gamma$, i.e. $\{\vec{\alpha}(0),\vec{\beta}(0)\}$
and $\{\vec{\alpha}(T), \vec{\beta}(T)\}$. As the initially prepared $W$ state is an eigenstate of $S_3 + T_3$ and 
$\mid\vec{\alpha}(t)\mid= 
\mid\vec{\beta}(t)\mid = \pi$ we can choose $\alpha_3 (0)= \beta_3(0) = \pi$.
Alternatively also the choice $\alpha_3 (0)= \beta_3(0) = -\pi$  would have been possible.
The final values of the curve $\gamma$ at $t=T$ can be determined from the transformation properties explicitly worked out 
in Appendix~\ref{app:rotSTbasis}. For $\mid\vec{\alpha}(t)\mid= \mid\vec{\beta}(t)\mid = \pi$ they reduce to the conditions
\begin{eqnarray}
\frac{\pi^2}{\sqrt{2}}&=& \mid \alpha_3(T)\beta_2(T)+\alpha_2(T)\beta_3(T)\mid,\nonumber\\
0 &=& \alpha_1(T)\beta_1(T)+\alpha_2(T)\beta_2(T)-\alpha_3(T)\beta_3(T),\nonumber \\
0 &=& \alpha_1(T)\beta_3(T)+\alpha_3(T)\beta_1(T),\nonumber \\
\frac{\pi^2}{\sqrt{2}}&=& \mid \alpha_2(T)\beta_1(T)-\alpha_1(T)\beta_2(T)\mid.
\label{GHZ-spherical}
\end{eqnarray}
Numerical solutions of these equations, which are consistent with the anholonomic boundary condition of Eq.~\eqref{anholexplicit},
are presented in Table I of Appendix~\ref{app:condGHZ}. For completeness, let us mention that the inverse conversion from 
a GHZ- to a $W$ state is easily realized through a time-reversal of the curve $\gamma$. Such a time reversal 
formally corresponds to interchanging $t=0$ and $t=T$ in Eq.~\eqref{anholexplicit} and can practically be achieved
using Rabi frequencies of the same magnitude as those in Eq.~\eqref{eq:omegapi}, but with the opposite sign.

Depending on the choice of the function $\dot{\theta}_{\alpha}(t)$ in Eq.~\eqref{anholexplicit} different time-dependent 
Rabi frequencies of the Hamiltonian~\eqref{eq:HeffZheng} can enable the desired quantum state conversion from a $W$ to a 
GHZ state. One of the simplest choices for $\dot{\theta}_{\alpha}(t)$ is a constant function which yields
\begin{eqnarray} \label{eq:theta'const}
\theta_{\alpha}(t) &=& t\:\frac{\theta_{\alpha}(T)}{T},~~t\in[0,T] \:, 
\end{eqnarray}
where in writing the last equation use has been made of the fact that $\theta_{\alpha}(0)=0$.
According to Eqs.~\eqref{eq:omegapi} it gives rise to time independent Rabi frequencies corresponding to an 
instantaneous turn on and turn off of the laser pulses inducing these Rabi frequencies at $t=0$ and $t=T$. Another 
possible choice for $\dot{\theta}_{\alpha}(t)$ is 
\begin{eqnarray}\label{eq:thetaswitch}
\dot{\theta}_{\alpha}(t) &=& \frac{\theta_\alpha(T)}{T(1-\tau)}
\left\{
\begin{array}{ll}
	 t/(\tau T),& 0 \leq t \leq \tau T\\
	 1, & \tau T\leq t \leq T-\tau T \\
	 (T-t)/(\tau T), & T - \tau T \leq t \leq T
\end{array}
\right.\nonumber\\
\end{eqnarray}
which yields (continuous) Rabi frequencies vanishing at $t=0$ and at $t=T$ and being turned on and off during a time 
interval of duration $\tau T$. 

\begin{figure}[t!]
	\includegraphics[width=8.6cm]{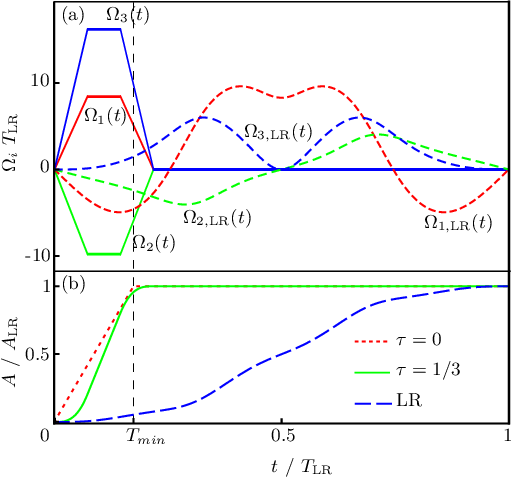} 
	\caption{(Color online)(a) Time evolution of Rabi frequencies for converting a $W$ to a GHZ state, compared to 
	the pulse sequences of the LR scheme of Ref.~\cite{Zheng+:20}.\\
	(b) Total squared pulse areas $A(t)$ [cf. Eq.~\eqref{eq:SqPA}] for different pulse sequences: $T_{\textrm{LR}}$ 
	is the time required for the desired 
	$W$-to-GHZ state conversion in the LR scheme, while $T_{min}$ is the corresponding time in the case of time-independent 
	Rabi frequencies ($\tau=0$).}
	\label{fig:pulses}
\end{figure}

\begin{figure}[b!]
	\includegraphics[width=0.95\linewidth]{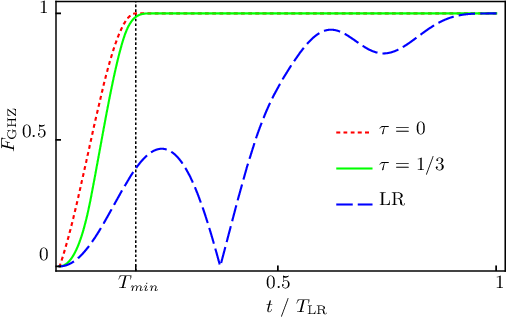}
	\caption{(Color online) Time evolution of the GHZ-state fidelity $F_{\textrm{GHZ}}=|\braket{\textrm{GHZ}}{\psi(t)}|$
		for the pulse sequences displayed in Fig.~\ref{fig:pulses}.}
	\label{fig:fidelities}
\end{figure}

In Figs.~\ref{fig:pulses} and \ref{fig:fidelities} characteristic features of the quantum state conversion from a $W$ to a GHZ 
state resulting from these two types of time dependent Rabi frequencies are compared with the corresponding results of the recently 
proposed STA-based scheme of Ref.~\cite{Zheng+:20} which is based on the Lewis-Riesenfeld (LR) theory. For this comparison the Rabi 
frequencies of these different quantum state conversion schemes are normalized such that their total squared pulse areas, i.e.
\begin{equation}
A(t) = \int_0^t \sum_{i=3}^3\Omega_i^2(t')  dt'\:, \label{eq:SqPA}
\end{equation}
are equal for times $t$ that correspond to the total
pulse durations ($t=T_{\rm min}$ and $t=T_{\rm LR}$, respectively). This implies that in all these schemes the same (time averaged) 
laser energy is required for achieving the 
quantum state conversion. In particular, we elect to normalize them to the total squared pulse area $A_{\text{LR}}$ 
corresponding to the LR scheme [cf. Fig.~\ref{fig:pulses}(b)].

The time evolution of the three Rabi frequencies capable of converting a $W$ state to a GHZ state according to Eq.~\eqref{eq:thetaswitch}
with  parameters $\theta_{\alpha}(T) = 1.92423$, $\theta_{\beta}(T) =0.906373$, $q_1 = q_3=1$, $q_2 =-1$ are depicted by the solid lines in 
Fig.~\ref{fig:pulses}(a). In this example the relative turn-on and turn-off time $\tau$ [cf. Eq.~\eqref{eq:thetaswitch}] of these time-dependent 
Rabi frequencies has been chosen as one third of the total pulse duration (i.e. $\tau=1/3$). These time-dependent Rabi frequencies are compared 
with the recently proposed pulse sequences resulting from a LR invariant~\cite{Zheng+:20} (dashed lines). 
All Rabi frequencies, in both our approach and in that of Ref.~\cite{Zheng+:20}, are assumed to vanish outside of their 
respective time interval $[0,T]$. Time is plotted in units of the time $T_{\textrm{LR}}$ which is required to complete the state 
conversion $\ket{W}\to\ket{\rm GHZ}$ in the LR scheme. 
It is apparent from Fig.~\ref{fig:fidelities} that the time-dependent Rabi frequencies resulting from Eq.~\eqref{eq:thetaswitch} 
are capable of completing 
the quantum state conversion in a significantly shorter time close to $T_{\rm min}$ than the ones based on the LR scheme. 
The minimal conversion time $T_{\rm min}$ is achieved with time-independent Rabi frequencies ($\tau = 0$) described by 
Eq.~\eqref{eq:theta'const}.
The time evolutions of the GHZ-state fidelites $F_{\textrm{GHZ}}=|\braket{\textrm{GHZ}}{\psi(t)}|$ corresponding to 
the pulse sequences shown 
in Fig.~\ref{fig:pulses} are depicted in Fig.~\ref{fig:fidelities}. Again the total squared pulse areas are equal for all three cases shown, 
namely 
the case of constant Rabi frequencies ($\tau=0$), Rabi frequencies turned on and off during one third of the pulse duration ($\tau=1/3$), 
and the Rabi frequencies of the LR scheme from Ref.~\cite{Zheng+:20}. 

In the system at hand -- as usual for optical experiments in general -- deleterious 
effects of decoherence are unavoidable. For instance, laser manipulation of atomic states leads
to phases that depend on atomic positions. Therefore, atomic motion inevitably
gives rise to dephasing. Another important effect in the presence of laser excitation
is spontaneous emission, which represents a major relaxation channel in the system under
consideration. In what follows, we briefly discuss these decoherence processes within
the framework of a Lindblad master equation~\cite{BreuerPetruccioneBOOK:02}. We show that -- for 
realistic values of the parameters that describe them -- such imperfections have only a small 
effect on the fidelities of GHZ states resulting from our proposed state-conversion protocol.

In order to describe spontaneous emission originating from radiative decay, for example, the 
Lindblad operator
\begin{equation}
L_{\rm Sp } = \sqrt{\Gamma}\sum_{n=1}^N\ket{g}_{nn}\bra{r}
\end{equation}
is used. It describes the spontaneous decay of individual atoms from the Rydberg- 
to the ground state, with $\Gamma$ being the corresponding decay rate. Analogously, 
the operator 
\begin{equation}
L_{\rm De } = \sqrt{\gamma}\sum_{n=1}^N\left(\ket{g}_{nn}\bra{g}-\ket{r}_{nn}\bra{r}\right)
\end{equation}
describes dephasing of Rydberg- and ground states with the dephasing rate $\gamma$.
Using the full interaction Hamiltonian of the system [cf. Eq.~\eqref{eq:Hint}] and taking 
$\rho(0)=\ket{W}\bra{W}$ as the initial state of the system  at $t=0$, we solve the corresponding 
Lindblad master equation numerically~\cite{Johansson+:12+13} and evaluate the resulting (open-system) 
fidelity with respect to the GHZ state
\begin{equation}
\ket{\textrm{GHZ}}_{\rm Int} =\frac{1}{\sqrt{2}}\left(\ket{ggg}+e^{i(\varphi-\Delta{E} t/\hbar)}
\ket{rrr}\right)   
\label{GHZnew}
\end{equation}
with
\begin{equation}
\Delta E/\hbar = 3 V - \left(\frac{3\Omega_{r0}^2}{\Delta_0}+\frac{3\Omega_{r0}^2}{\Delta_0-2V}\right) \ .
\end{equation}
The time-dependent phase involving
$\Delta E$ compensates for the fact that the GHZ state defined in Eq.~\eqref{eq:psit} and the effective 
Hamiltonian of Eq.~\eqref{eq:HeffZheng} refer to an interaction picture in which the energy shifts originating 
from the interatomic interaction and from the strong laser field with Rabi frequency $\Omega_{r0}$ have been 
taken into account whereas the Hamiltonian of Eq.~\eqref{eq:Hint} and the GHZ-state of Eq.~\eqref{GHZnew} refer 
to an interaction picture without these energy shifts.

\begin{figure}[t!]
\includegraphics[width=0.95\linewidth]{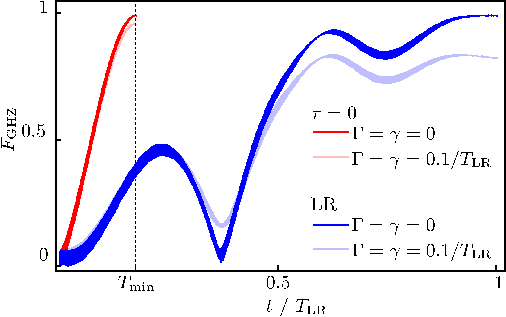}
\caption{(Color online) Time evolution of the GHZ-state fidelity ${F}_\text{GHZ}$ in the open system 
approach for our time-independent pulse sequences and for the pulse sequences of the LR scheme of Ref.~\cite{Zheng+:20}.}
\label{fig:fidelities_diss}
\end{figure}

To demonstrate the robustness of our state-conversion protocol against decoherence (for an illustration, see 
Fig.~\ref{fig:fidelities_diss}), we investigate the open-system dynamics for the case of rather strong decoherence and 
dissipation with parameters $\Gamma = \gamma = 0.1/T_{\rm LR}$, $T_{\rm LR}\Omega_{r0} = 500$, $T_{\rm LR}V= 
2\pi\times 1000$, and $T_{\rm LR}\Delta_0= -2\pi\times 3000$. Figure~\ref{fig:fidelities_diss} compares the obtained GHZ-state 
fidelities for our time-independent scheme ($\tau = 0$) and the one of Ref.~\cite{Zheng+:20} (LR), showing at the same time the 
results obtained in the closed-system scenario ($\Gamma=\gamma=0$) for both of these schemes. Because our scheme allows one to 
carry out the desired state conversion in a significantly shorter time, its corresponding GHZ-state fidelity is much less affected 
by decoherence, as its deleterious effects accumulate over time. Even with the above values for the decoherence and dissipation 
rates $\Gamma$ and $\gamma$, our scheme still preserves a fidelity that exceeds $96\%$ for $t\approx T_{\rm min}$. [Note that 
the broadening of the lines in Fig.~\ref{fig:fidelities_diss} is a consequence of small oscillations, 
which are not taken into account in the rotating-wave approximation used for the derivation of the effective Hamiltonian of 
Eq.~\eqref{eq:HeffZheng} and which could be minimized by choosing an even larger detuning $\Delta_0$.]

Owing to the advanced experimental capabilities currently available, even an application of complex time-dependent laser pulses 
of the kind obtained in Ref.~\cite{Zheng+:20} is in principle feasible. Yet, our scheme -- with its resulting time-independent 
Rabi frequencies -- is more easily applicable experimentally, especially from the standpoint of scaling to larger systems. As can 
be inferred from Fig.~\ref{fig:fidelities}, an important additional advantage of our scheme -- compared to that of 
Ref.~\cite{Zheng+:20} -- is that the time $T_{\rm min}$ it requires for the conversion of a $W$ state into its GHZ counterpart 
is significantly shorter than the corresponding time $T_{\textrm{LR}}$ in Ref.~\cite{Zheng+:20}.

\section{Summary and Conclusions} \label{SummConcl}
To summarize, in this paper we addressed the problem of $W$-to-GHZ state 
conversion in the Rydberg-blockade regime of a neutral atom system in which 
each atom is driven by four external laser pulses. While the same problem has 
recently been investigated using shortcuts to adiabaticity -- more precisely, 
techniques based on Lewis-Riesenfeld invariants~\cite{Zheng+:20} -- we have 
treated it using a completely different, Lie-algebraic approach based on the 
dynamical symmetries of the underlying system.

Because it employs Lie-algebraic techniques in the context of quantum-state control, 
our work presents a significant degree of novelty from the methodological standpoint. 
Namely, the use of Lie-algebraic concepts in the realm of quantum control has heretofore 
been almost exclusively confined to the realm of quantum operator 
control~\cite{Ramakrishna+Rabitz:96,Heule++:10} -- typically in the context of quantum-gate 
optimization~\cite{StojanovicToffoli:12,Zahedinejad+:15} -- where they provide the mathematical 
underpinnings of the concept of complete controllability~\cite{Ramakrishna+Rabitz:96,Stojanovic:19}. 

Most importantly, our resulting state-conversion protocol has two principal advantages 
compared to that of Ref.~\cite{Zheng+:20}. Firstly, unlike the latter work, which leads 
to a state-conversion protocol involving strongly time-dependent Rabi frequencies of 
external lasers, our approach results in a signficantly simpler one, which even allows for
time-independent Rabi frequencies. Secondly, our approach allows the sought-after 
$W$-to-GHZ state conversion to be carried out significantly faster than that of Ref.~\cite
{Zheng+:20}. Both of these advantages also speak in favour of an easier experimental 
implementation of our protocol, as well as its better scalability. 

With minor modifications, our approach can be generalized, not only to other state-conversion 
problems in the Rydberg-atom system under consideration, but also to systems belonging
to other QIP platforms. An experimental implementation of our state-conversion protocol
is keenly anticipated.
\begin{acknowledgments}
The authors acknowledge useful discussions with G. Birkl and A. R. P. Rau. This research was 
supported by the Deutsche Forschungsgemeinschaft (DFG) -- SFB 1119 -- 236615297.
\end{acknowledgments}

\appendix
\section{Unitary transformations of the pseudospin states in Eqs.~(\ref{upup}) and (\ref{uprest})}\label{app:rotSTbasis}
We consider a general unitary transformation as defined by Eq.~\eqref{unitary}.
From the four-dimensional representation of the angular-momentum operators $S_i$ and $T_i$ [cf. 
Eq.~\eqref{eq:Liealgebrarepresentation}], it is straightforward to obtain the relations
\begin{eqnarray}
e^{-i\vec{\alpha}\cdot\vec{S}}&=&\cos\frac{|\vec{\alpha}|}{2}-2i\sin\frac{\vec\alpha\cdot\vec{S}}
{|\vec{\alpha}|}\sin\frac{|\vec{\alpha}|}{2},\nonumber\\
e^{-i\vec{\beta}\cdot\vec{T}}&=&\cos\frac{|\vec{\beta}|}{2}-2i\sin\frac{\vec\beta\cdot\vec{T}}
{|\vec{\beta}|}\sin\frac{|\vec{\beta}|}{2}.
\label{Uexplicit}
\end{eqnarray}
For the pseudospin states of Eqs.~\eqref{upup}) and \eqref{uprest} the last two relations imply
that 
\begin{eqnarray}
e^{-i\vec{\alpha}\cdot\vec{S}}\ket{\uparrow \nu} &=&
M_{++}(\vec\alpha) \ket{\uparrow \nu} + 
M_{+-}(\vec\alpha) \ket{\downarrow \nu}\:,\nonumber\\
e^{-i\vec{\alpha}\cdot\vec{S}}\ket{\downarrow \nu} &=&
-M^*_{+-}(\vec\alpha) \ket{\uparrow \nu} + 
M^*_{++}(\vec\alpha) \ket{\downarrow \nu} \:,
\end{eqnarray}
where $\nu \in \{\uparrow,\downarrow\}$ and 
\begin{eqnarray}
M_{++}(\vec{\alpha}) &=& \cos\frac{|\vec{\alpha}|}{2}
 -i\frac{\alpha_3}{|\vec{\alpha}|}\sin\frac{|\vec{\alpha}|}{2},\nonumber\\
M_{+-}(\vec{\alpha}) &=&  \frac{(-i \alpha_1+\alpha_2)}{|\vec{\alpha}|}\sin\frac{|\vec{\alpha}|}{2}. 
\end{eqnarray}
In deriving the last results, use has been made of the fact that the angular-momentum operators $S_i$ act on 
the first pseudospin only. Because $|M_{++}(\vec{\alpha})|^2+|M_{+-}(\vec{\alpha})|^2=1$, 
for $\vec{\alpha}\in {\mathbb R}^3$, these unitary transformations belong to the group $SL(2, {\mathbb C})$.

By applying an additional unitary transformation generated by the angular-momentum operators $T_j$ -- which 
commute with $S_i$ ($i,j\in \{1,2,3\}$) and act on the second pseudospin only -- we finally obtain the following 
relations:
\begin{align}
&e^{-i\vec{\alpha}\cdot \vec{S}}e^{-i\vec{\beta}\cdot \vec{T}} \ket{\uparrow \uparrow}~\longrightarrow~ \nonumber \\
&\frac{1}{\sqrt{2}}
\left(
\begin{array}{c}
-iM_{++}(\vec{\alpha}) M_{++}(\vec{\beta})-iM_{+-}(\vec{\alpha}) M_{+-}(\vec{\beta})\\
-iM_{++}(\vec{\alpha}) M_{+-}(\vec{\beta})-iM_{+-}(\vec{\alpha}) M_{++}(\vec{\beta})\\
M_{++}(\vec{\alpha}) M_{++}(\vec{\beta})-M_{+-}(\vec{\alpha}) M_{+-}(\vec{\beta})\\
-M_{++}(\vec{\alpha}) M_{+-}(\vec{\beta})+M_{+-}(\vec{\alpha}) M_{++}(\vec{\beta})
\end{array}
\right),\nonumber\\
&e^{-i\vec{\alpha}\cdot \vec{S}}e^{-i\vec{\beta}\cdot \vec{T}}\ket{\downarrow\uparrow}~\longrightarrow~\nonumber \\
&\frac{1}{\sqrt{2}}
\left(
\begin{array}{c}
-iM^{*}_{++}(\vec{\alpha}) M_{+-}(\vec{\beta})+iM^{*}_{+-}(\vec{\alpha}) M_{++}(\vec{\beta})\\
-iM^{*}_{++}(\vec{\alpha}) M_{++}(\vec{\beta})+iM^{*}_{+-}(\vec{\alpha}) M_{+-}(\vec{\beta})\\
-M^{*}_{++}(\vec{\alpha}) M_{+-}(\vec{\beta})-M^{*}_{+-}(\vec{\alpha}) M_{++}(\vec{\beta})\\
M^{*}_{++}(\vec{\alpha}) M_{++}(\vec{\beta})+M^{*}_{+-}(\vec{\alpha}) M_{+-}(\vec{\beta})
\end{array}
\right),\nonumber\\
&e^{-i\vec{\alpha}\cdot \vec{S}}e^{-i\vec{\beta}\cdot \vec{T}}\ket{\uparrow \downarrow}~\longrightarrow~ \nonumber \\
&\frac{1}{\sqrt{2}}
\left(
\begin{array}{c}
iM_{++}(\vec{\alpha}) M^{*}_{+-}(\vec{\beta})-iM_{+-}(\vec{\alpha}) M^{*}_{++}(\vec{\beta})\\
-iM_{++}(\vec{\alpha}) M^{*}_{++}(\vec{\beta})+iM_{+-}(\vec{\alpha}) M^{*}_{+-}(\vec{\beta})\\
-M_{++}(\vec{\alpha}) M^{*}_{+-}(\vec{\beta})-M_{+-}(\vec{\alpha}) M^{*}_{++}(\vec{\beta})\\
-M_{++}(\vec{\alpha}) M^{*}_{++}(\vec{\beta})-M_{+-}(\vec{\alpha}) M^{*}_{+-}(\vec{\beta})
\end{array}
\right),\nonumber\\
&e^{-i\vec{\alpha}\cdot \vec{S}}e^{-i\vec{\beta}\cdot \vec{T}}\ket{\downarrow \downarrow}~\longrightarrow~ \nonumber \\
&\frac{1}{\sqrt{2}}
\left(
\begin{array}{c}
-iM^{*}_{++}(\vec{\alpha}) M^{*}_{++}(\vec{\beta})-iM^{*}_{+-}(\vec{\alpha}) M^{*}_{+-}(\vec{\beta})\\
+iM^{*}_{++}(\vec{\alpha}) M^{*}_{+-}(\vec{\beta})+iM^{*}_{+-}(\vec{\alpha}) M^{*}_{++}(\vec{\beta})\\
-M^{*}_{++}(\vec{\alpha}) M^{*}_{++}(\vec{\beta})+M^{*}_{+-}(\vec{\alpha}) M^{*}_{+-}(\vec{\beta})\\
-M^{*}_{++}(\vec{\alpha}) M^{*}_{+-}(\vec{\beta})+M^{*}_{+-}(\vec{\alpha}) M^{*}_{++}(\vec{\beta})
\end{array}
\right).
\label{eq:rotatedeigenvectors}
\end{align}

\section{Parameters describing GHZ states}\label{app:condGHZ}
\begin{table}[t!]
	\begin{tabular}[t]{|c|c|c|c|c|c|c|}
		\hline
		$\theta_{\alpha}(T)$ &$\theta_{\beta}(T)$ &$\phi_{\alpha}(0)$& $\phi_{\beta}(0)$& $q_1$ &  $q_2$ & $q_3$\\
		\hline
		$1.92423$&$0.906373$&$4.33454$&$2.47062$&$+1$&$-1$&$+1$\\
		\hline
		$1.92423$&$0.906373$&$1.94864$&$3.81256$&$+1$&$+1$&$+1$\\
		\hline
		$1.92423$&$0.906373$&$1.19295$&$5.61221$&$-1$&$+1$&$+1$\\
		\hline
		$1.92423$&$0.906373$&$5.09024$&$0.670972$&$-1$&$-1$&$+1$\\
		\hline
		$0.906373$&$1.92423$&$2.47062$&$4.33454$&$-1$&$+1$&$-1$\\
		\hline
		$0.906373$&$1.92423$&$3.81256$&$1.94864$&$-1$&$-1$&$-1$\\
		\hline
		$0.906373$&$1.92423$&$5.61221$&$1.19295$&$-1$&$-1$&$-1$\\
		\hline
		$0.906373$&$1.92423$&$0.670972$&$5.09024$&$+1$&$+1$&$-1$\\
		\hline
	\end{tabular}
	\caption{Spherical coordinates in the parameter space of the Lie algebra $su(2)\oplus su(2)$ 
	describing GHZ states as defined by Eq.(\ref{eq:psit}).} \label{tab:coordqi}
\end{table}
In order to determine the parameters $\{\vec{\alpha}(T),\vec{\beta}(T)\}$ which describe a GHZ state [as defined 
by Eq.~\eqref{eq:psit}], we start from its expansion in terms of the orthonormal basis states of Eqs.~\eqref{upup}
and \eqref{uprest}, i.e.
\begin{eqnarray}
\ket{\textrm{GHZ}} &=& \frac{1}{2}\left(i\ket{\uparrow\uparrow} + i \ket{\downarrow\downarrow}
+e^{i\varphi}\ket{\downarrow\uparrow}- e^{i\varphi}\ket{\uparrow\downarrow} \right).\nonumber\\
\end{eqnarray}
With the aid of Eqs.~\eqref{eq:rotatedeigenvectors} we obtain the relations
\begin{eqnarray}
\pm\frac{1}{\sqrt{2}}&=&-\Im\left(M_{++}(\vec{\alpha}(T))M^*_{+-}(\vec{\beta}(T))\right)+\nonumber\\
&&
\Im\left(M_{+-}(\vec{\alpha}(T))M^*_{++}(\vec{\beta}(T))\right),\nonumber\\ 
0 &=& \Re\left(M_{++}(\vec{\alpha}(T))M_{++}^*(\vec{\beta}(T))\right) -\nonumber\\
&&\Re\left(M_{+-}(\vec{\alpha}(T))M_{+-}^*(\vec{\beta}(T))\right),\nonumber\\
0 &=& \Re\left(M_{++}(\vec{\alpha}(T))M_{+-}^*(\vec{\beta}(T))\right)+\nonumber\\
&&\Re\left(M_{+-}(\vec{\alpha}(T))M_{++}^*(\vec{\beta}(T))\right),\nonumber\\
\pm \frac{1}{\sqrt{2}}&=&\Im\left(M_{++}(\vec{\alpha}(T))M^*_{++}(\vec{\beta}(T))\right)+\nonumber\\
&&\Im\left(M_{+-}(\vec{\alpha}(T))M^*_{+-}(\vec{\beta}(T))\right) \label{eq:condGHZ} 
\end{eqnarray}
for the parameters of a GHZ state, where $\Re$ ($\Im$) denotes the real (imaginary) part of a complex quantity.

In the special case of $\mid\vec{\alpha}(T)\mid = \mid\vec{\beta}(T)\mid = \pi$ these relations reduce to 
Eqs.~\eqref{GHZ-spherical}), whose general solutions are given by
\begin{eqnarray}
\vec{\alpha}(T) &\longrightarrow&
\left(
\begin{array}{c}
q_1\alpha_3\\
q_2 \sqrt{\pi^2 - 2\alpha_3^2}\\
\alpha_3
\end{array}
\right),\nonumber\\
\vec{\beta}(T) &\longrightarrow&
\left(
\begin{array}{c}
-q_1q_3\sqrt{\pi^2 - 2\alpha_3^2}\\
2q_2 q_3 \alpha_3\\
q_3 \sqrt{\pi^2 - 2\alpha_3^2}
\end{array}
\right)
\end{eqnarray}
with $-\pi/\sqrt{2}\leq \alpha_3 \leq \pi/\sqrt{2}$ and $q_1,q_2,q_3 \in \{-1,+1\}$.
Introducing spherical coordinates yields the relations
\begin{eqnarray}
\cos\phi_{\alpha}(0) &=&q_1 \cot\theta_{\alpha}(T),\nonumber\\
\cos\phi_{\beta}(0) &=&-q_1 \cot\theta_{\beta}(T),\nonumber\\
\cos^2\theta_{\alpha}(T)+ \cos^2 \theta_{\beta}(T) &=&\frac{1}{2}
\end{eqnarray}
which have to be fulfilled necessarily by the spherical coordinates of a GHZ state.
In addition, these coordinates have to fulfill the anholonomic boundary conditions of Eq.(\ref{anholexplicit}) 
relating $\theta_{\beta}(T)$ and $\theta_{\alpha}(T)$. This leads to a unique set of spherical 
coordinates for each combination of $q_1$, $q_2$, $q_3$ describing the desired state conversion. These
sets are given explicitly in Table~\ref{tab:coordqi}.

\end{document}